\documentclass[aps,prd,reprint,showpacs,showkeys,nofootinbib,superscriptaddress]{revtex4-1}

\usepackage{amsmath}
\usepackage{graphicx}

\begin{document}

\title{Searches for electron interactions induced by new physics in the EDELWEISS-III germanium bolometers}

\author{E.~Armengaud}
\email{eric.armengaud@cea.fr}
\affiliation{IRFU, CEA, Universit\'{e} Paris-Saclay, F-91191 Gif-sur-Yvette, France}
\author{C.~Augier}
\affiliation{Univ Lyon, Universit\'{e} Lyon 1, CNRS/IN2P3, IPN-Lyon, F-69622, Villeurbanne, France}
\author{A.~Beno\^{i}t}
\affiliation{Institut N\'{e}el, CNRS/UJF, 25 rue des Martyrs, BP 166, 38042 Grenoble, France}
\author{L.~Berg\'{e}}
\affiliation{CSNSM, Universit\'{e} Paris-Sud, CNRS/IN2P3, Universit\'{e} Paris-Saclay, 91405 Orsay, France}
\author{J.~Billard}
\affiliation{Univ Lyon, Universit\'{e} Lyon 1, CNRS/IN2P3, IPN-Lyon, F-69622, Villeurbanne, France}
\author{A.~Broniatowski}
\affiliation{CSNSM, Universit\'{e} Paris-Sud, CNRS/IN2P3, Universit\'{e} Paris-Saclay, 91405 Orsay, France}
\author{P.~Camus}
\affiliation{Institut N\'{e}el, CNRS/UJF, 25 rue des Martyrs, BP 166, 38042 Grenoble, France}
\author{A.~Cazes}
\affiliation{Univ Lyon, Universit\'{e} Lyon 1, CNRS/IN2P3, IPN-Lyon, F-69622, Villeurbanne, France}
\author{M.~Chapellier}
\affiliation{CSNSM, Universit\'{e} Paris-Sud, CNRS/IN2P3, Universit\'{e} Paris-Saclay, 91405 Orsay, France}
\author{F.~Charlieux}
\affiliation{Univ Lyon, Universit\'{e} Lyon 1, CNRS/IN2P3, IPN-Lyon, F-69622, Villeurbanne, France}
\author{M. De~J\'{e}sus}
\affiliation{Univ Lyon, Universit\'{e} Lyon 1, CNRS/IN2P3, IPN-Lyon, F-69622, Villeurbanne, France}
\author{L.~Dumoulin}
\affiliation{CSNSM, Universit\'{e} Paris-Sud, CNRS/IN2P3, Universit\'{e} Paris-Saclay, 91405 Orsay, France}
\author{K.~Eitel}
\affiliation{Karlsruher Institut f\"{u}r Technologie, Institut f\"{u}r Kernphysik, Postfach 3640, 76021 Karlsruhe, Germany}
\author{J.~Gascon}
\affiliation{Univ Lyon, Universit\'{e} Lyon 1, CNRS/IN2P3, IPN-Lyon, F-69622, Villeurbanne, France}
\author{A.~Giuliani}
\affiliation{CSNSM, Universit\'{e} Paris-Sud, CNRS/IN2P3, Universit\'{e} Paris-Saclay, 91405 Orsay, France}
\author{M.~Gros}
\affiliation{IRFU, CEA, Universit\'{e} Paris-Saclay, F-91191 Gif-sur-Yvette, France}
\author{Y.~Jin}
\affiliation{Laboratoire de Photonique et de Nanostructures, CNRS, Route de Nozay, 91460 Marcoussis, France}
\author{A.~Juillard}
\affiliation{Univ Lyon, Universit\'{e} Lyon 1, CNRS/IN2P3, IPN-Lyon, F-69622, Villeurbanne, France}
\author{M.~Kleifges}
\affiliation{Karlsruher Institut f\"{u}r Technologie, Institut f\"{u}r Prozessdatenverarbeitung und Elektronik, Postfach 3640, 76021 Karlsruhe, Germany}
\author{V.~Kozlov}
\altaffiliation[Present address: ]{Karlsruher Institut f\"{u}r Technologie, Steinbuch Centre for Computing, 76128 Karlsruhe, Germany}
\affiliation{Karlsruher Institut f\"{u}r Technologie, Institut f\"{u}r Experimentelle Teilchenphysik, Gaedestr. 1, 76128 Karlsruhe, Germany}
\author{H.~Kraus}
\affiliation{University of Oxford, Department of Physics, Keble Road, Oxford OX1 3RH, UK}
\author{V. A.~Kudryavtsev}
\affiliation{University of Sheffield, Department of Physics and Astronomy, Sheffield, S3 7RH, UK}
\author{H.~Le-Sueur}
\affiliation{CSNSM, Universit\'{e} Paris-Sud, CNRS/IN2P3, Universit\'{e} Paris-Saclay, 91405 Orsay, France}
\author{R.~Maisonobe}
\affiliation{Univ Lyon, Universit\'{e} Lyon 1, CNRS/IN2P3, IPN-Lyon, F-69622, Villeurbanne, France}
\author{S.~Marnieros}
\affiliation{CSNSM, Universit\'{e} Paris-Sud, CNRS/IN2P3, Universit\'{e} Paris-Saclay, 91405 Orsay, France}
\author{D.~Misiak}
\affiliation{Univ Lyon, Universit\'{e} Lyon 1, CNRS/IN2P3, IPN-Lyon, F-69622, Villeurbanne, France}
\author{X.-F.~Navick}
\affiliation{IRFU, CEA, Universit\'{e} Paris-Saclay, F-91191 Gif-sur-Yvette, France}
\author{C.~Nones}
\affiliation{IRFU, CEA, Universit\'{e} Paris-Saclay, F-91191 Gif-sur-Yvette, France}
\author{E.~Olivieri}
\affiliation{CSNSM, Universit\'{e} Paris-Sud, CNRS/IN2P3, Universit\'{e} Paris-Saclay, 91405 Orsay, France}
\author{P.~Pari}
\affiliation{IRAMIS, CEA, Universit\'{e} Paris-Saclay, F-91191 Gif-sur-Yvette, France}
\author{B.~Paul}
\affiliation{IRFU, CEA, Universit\'{e} Paris-Saclay, F-91191 Gif-sur-Yvette, France}
\author{D.~Poda}
\affiliation{CSNSM, Universit\'{e} Paris-Sud, CNRS/IN2P3, Universit\'{e} Paris-Saclay, 91405 Orsay, France}
\author{E.~Queguiner}
\affiliation{Univ Lyon, Universit\'{e} Lyon 1, CNRS/IN2P3, IPN-Lyon, F-69622, Villeurbanne, France}
\author{S.~Rozov}
\affiliation{JINR, Laboratory of Nuclear Problems, Joliot-Curie 6, 141980 Dubna, Moscow Region, Russian Federation}
\author{V.~Sanglard}
\affiliation{Univ Lyon, Universit\'{e} Lyon 1, CNRS/IN2P3, IPN-Lyon, F-69622, Villeurbanne, France}
\author{S.~Scorza}
\altaffiliation[Present address: ]{SNOLAB, Lively, Ontario, Canada.}
\affiliation{Karlsruher Institut f\"{u}r Technologie, Institut f\"{u}r Experimentelle Teilchenphysik, Gaedestr. 1, 76128 Karlsruhe, Germany}
\author{B.~Siebenborn}
\affiliation{Karlsruher Institut f\"{u}r Technologie, Institut f\"{u}r Kernphysik, Postfach 3640, 76021 Karlsruhe, Germany}
\author{D.~Tcherniakhovski}
\affiliation{Karlsruher Institut f\"{u}r Technologie, Institut f\"{u}r Prozessdatenverarbeitung und Elektronik, Postfach 3640, 76021 Karlsruhe, Germany}
\author{L.~Vagneron}
\affiliation{Univ Lyon, Universit\'{e} Lyon 1, CNRS/IN2P3, IPN-Lyon, F-69622, Villeurbanne, France}
\author{M.~Weber}
\affiliation{Karlsruher Institut f\"{u}r Technologie, Institut f\"{u}r Prozessdatenverarbeitung und Elektronik, Postfach 3640, 76021 Karlsruhe, Germany}
\author{E.~Yakushev}
\affiliation{JINR, Laboratory of Nuclear Problems, Joliot-Curie 6, 141980 Dubna, Moscow Region, Russian Federation}
\author{A.~Zolotarova}
\affiliation{IRFU, CEA, Universit\'{e} Paris-Saclay, F-91191 Gif-sur-Yvette, France}

\collaboration{EDELWEISS Collaboration}
\noaffiliation

\begin{abstract}
We make use of the EDELWEISS-III array of germanium bolometers to search for electron interactions at the keV scale induced by  phenomena beyond the Standard Model. A 90\% C.L. lower limit is set on the electron lifetime decaying to invisibles, $\tau > 1.2\times 10^{24}$~years. We investigate the emission of axions or axionlike particles (ALPs) by the Sun, constraining the coupling parameters $g_{ae}<1.1\times 10^{-11}$ and $g_{ae}\times g_{aN}^{\rm eff} < 3.5\times 10^{-17}$ at 90\% C.L. in the massless limit. We also directly search for the absorption of bosonic dark matter particles that would constitute our local galactic halo. Limits are placed on the couplings of ALPs or hidden photon dark matter in the mass range 0.8 -- 500~keV/c$^2$. Prospects for searching for dark matter particles with masses down to 150~eV/c$^2$ using improved detectors are presented.
\end{abstract}


\maketitle

\section{Introduction}

The EDELWEISS-III experiment, located in the Laboratoire Souterrain de Modane (LSM), uses an array of 870 g germanium detectors. Their instrumentation is primarily optimized to identify hypothetical keV-scale energy depositions from nuclear recoils produced by elastic scattering of weakly interacting massive particles (WIMPs) that could constitute our galactic dark matter halo~\cite{Armengaud:2016cvl,Hehn:2016nll}.

In addition, the experiment is also able to detect electronic interactions generated by rare processes with energy depositions down to keV-scale energies. The associated sensitivity to such rare processes profits from the large recorded exposure, the overall low-background environment and the specific detector technology. In particular, the so-called fully interdigit (FID) electrode scheme allows the identification of interactions inside the fiducial volume of individual detectors and therefore suppresses backgrounds from surface radioactivity down to the experimental energy threshold. A measurement of the remaining fiducial electron recoil background associated with radioactive processes was presented and interpreted in~\cite{Armengaud:2016aoz}. In this article we present the results of searches for rare processes induced by several hypothetical phenomena, producing electron recoils in the 0.8 -- 500~keV energy range.

Most search channels focus on axions or axionlike particles (ALPs) emitted by the Sun or which would constitute the local dark matter halo. We also present a search for germanium K-shell electrons decaying into invisible particles. Our results improve significantly over the axion searches carried out with the EDELWEISS-II detectors~\cite{Armengaud2013b}. Other experiments also recently published similar searches, using either liquid Xenon targets~\cite{Aprile2014,XENONcollaboration2017,Abe2012a,Abe2014,Akerib2017,Fu2017} or crystalline detectors~\cite{Abgrall2016,Liu2017,Aalseth:2008rx,Angloher:2016rji}.

\subsection{Solar axions}

The first category of our search focuses on solar axions. QCD axions provide an elegant explanation for the observed lack of \textit{CP} violation in the sector of strong interactions. These pseudoscalar particles may have arbitrary masses $m_a$, whose value is dictated by the Peccei-Quinn energy scale $f_a$. Their couplings to ordinary matter, namely photons, electrons and nucleons, depend on this energy scale as well as on the exact realization of the axion model. These interactions can be represented by the following effective Lagrangian:

\begin{equation}
{\cal L} = -\frac{1}{4} g_{a\gamma} a F \tilde{F} + i g_{ae} \bar{e}\gamma_5 e \,a + i \bar{N}\gamma_5 (g_{aN}^0+g_{aN}^3\tau_3) N\, a.
\end{equation}

Here $a$ is the axion field, $F$ the photon, $e$ the electron and $N = (p , n)$ is the nucleon isospin doublet. The parameters $g_{aX}$ are then functions of $m_a$ for a given QCD axion model. They can, however, take any value in the case of a generic axionlike particle not related to the QCD sector. The Sun is a potential source of axions as long as the axion mass is smaller than its inner temperature. The solar axion flux is produced by different processes, each associated with the above-mentioned couplings. In this article we concentrate on two fluxes:
\begin{enumerate}
\item Thanks to their couplings to electrons, axions are produced via Compton-like, bremsstrahlung-like and recombination/de-excitationlike processes. The corresponding, so-called CBRD flux peaks around $1-2$~keV, and its intensity scales as $g_{ae}^2$. It has been calculated in~\cite{Redondo2013}.
\item $^{57}$Fe, largely present in the Sun's core, possesses a first excited state at $E^{\star}=14.4$~keV which may de-excite through the emission of an axion. The corresponding flux is monoenergetic, at 14.4~keV. Its line intensity scales with the combination $(g_{aN}^{\rm eff})^2 \equiv (-1.19\,g_{aN}^0+g_{aN}^3)^2$.
\end{enumerate}

\noindent Solar axions can be detected thanks to the equivalent of the photoelectric effect, i.e.\ the absorption of an axion by an electron, with the following cross section:

\begin{equation}
\sigma_{ae}(E) = \sigma_{pe}(E)\frac{3\,g_{ae}^2 E^2}{16\pi \alpha m_e^2 \beta} \left( 1-\frac{\beta^{2/3}}{3}\right).
\label{eqn:sigma_ae}
\end{equation}

Here $E$ is the total axion energy, identical to the electron recoil, and $\beta$ its velocity relative to the speed of light. $\sigma_{ae}$ is proportional to the photoelectric cross section in germanium $\sigma_{pe}$, as the same form factor associated to electronic wave functions is involved. In our calculations, we use NIST data~\cite{NIST}, complemented by~\cite{Veigele1973} for energies $E\le1.3$~keV.

The signals associated with both CBRD and $^{57}$Fe axions are proportional to the product of fluxes times $\sigma_{ae}$, convolved by the detector's energy resolution. While the $^{57}$Fe signal is an individual line at 14.4~keV, the energy distribution of the CBRD signal has a relatively broad shape in the energy range $1 \lesssim E \lesssim 4$~keV. The total intensities of these signals scale as $g_{ae}^4$ (CBRD) and $g_{ae}^2\times (g_{aN}^{\rm eff})^2$ ($^{57}$Fe), respectively.

\subsection{keV-scale bosonic dark matter}

While the dark matter (DM) paradigm is more and more strengthened by cosmological and astrophysical observations, the nature and, in particular, the mass of this physical object is completely unknown. The most studied thermal WIMP scenario naturally accomodates for masses in the GeV -- TeV range, but masses in the keV -- MeV range should also be explored~\cite{Adhikari:2016bei,Pospelov2008}. Thermal relics with a mass $\lesssim 4$~keV are severely constrained by observations of cosmological structures at small scales~\cite{Baur:2015jsy}, but much lower values for the mass of dark matter are possible if its relic density is driven by a nonthermal mechanism, as is the case for axions and ALPs.

Dark matter direct detection consists in searching for interactions of dark matter from our local galactic halo directly inside an experimental device. In the case of dark matter with a mass smaller than $\sim 10$~MeV/c$^2$, current direct detection experiments do not yet have the energy threshold to measure its elastic scattering on ordinary matter. However, if dark matter is made of bosons then its absorption can be measured down to much lower masses. Here we will search for the absorption of bosonic DM over a wide range of masses between 0.8 and 500~keV/c$^2$. Given the relatively well-known local DM mass density $\rho_{\rm DM} \simeq 0.3$~GeV/cm$^3$ and average velocity with respect to Earth, $\langle v\rangle \simeq 10^{-3} c$~\cite{Lewin:1995rx}, the DM flux reaching the detector is:

\begin{equation}
\Phi =  \frac{\rho_{\rm DM} \langle v\rangle}{m_a} \simeq 9.0\times 10^{12} \,\frac{\rm keV}{m_a} \; {\rm cm}^{-2}\cdot {\rm s}^{-1}.
\end{equation}

In order to quantify the intensity of the DM absorption process, we consider two simple cases. If DM is an ALP, then its coupling to electrons has the same expression as in the case of QCD axions, and the absorption cross section is identical to Eq.~\ref{eqn:sigma_ae}, with $\beta=10^{-3}$ and $E=m_a\,c^2$. If DM is a dark photon, labeled $A'$, we assume it is effectively coupled to the standard model at low energy through kinetic mixing with photons, ${\cal L} \sim \kappa F\,F'$. The absorption cross section for electrons is then simply proportional to the photoelectric cross section:

\begin{equation}
\sigma_{A'e}(m_{A'}) = \sigma_{pe}(E=m_{A'}\,c^2) \times \frac{\kappa^2 }{\beta}.
\end{equation}

The measurement of a monochromatic line of intensity $R$ at a given energy $E$ can then be interpreted as a DM absorption feature, with DM mass $m_a$ ($m_{A'}$) equal to $E$, and effective coupling $g_{ae}$ ($\kappa$) proportional to $\sqrt{R}$.

After extracting bounds on the absorption of DM with mass $> 0.8$~keV/c$^2$ using EDELWEISS-III data, we present future sensitivities to bosonic DM with a mass down to 150 eV/c$^2$ based on projections with detectors being currently developed.

\section{Experimental setup, data and analysis}\label{sec:setup}

\subsection{Setup}

The EDELWEISS-III infrastructure and related detector performance are described in~\cite{Armengaud2017c}. The experiment is located in the LSM, whose mean rock overburden of 4800 meters water-equivalent reduces the cosmic muon flux to 5~muons/m$^2\cdot$~day. The active target at the time of data taking relevant for this study consists of twenty-four germanium detectors cooled down to 18~mK and surrounded by a set of shields against radioactivity: namely 20~cm of lead against $\gamma$ rays and 50~cm of polyethylene against neutrons. A muon veto made of plastic scintillators surrounds the overall setup.

The detectors are 870 g, high-purity germanium cylindrical crystals, whose surfaces are all covered with ring-shaped, interleaved aluminium electrodes biased at alternate values of potentials. The resulting electric field separates the detector's volume in two parts. Electron-hole pairs produced inside the central, bulk volume are drifted by an axial electric field and collected by fiducial electrodes. Charges produced at a distance smaller than $\sim 2$~mm from the surfaces are drifted parallel to the surface and collected by fiducial and veto electrodes of a given side. In addition, the total phonon signal produced by an interaction is measured by two germanium neutron transmutation doped (NTD) thermistors. In the end, ionization signals permit the identification of fiducial interactions, and for those interactions the phonon signals provide an ionization yield measurement, and therefore permit the separation of electron recoils from nuclear recoils. Finally, the combination of both ionization and phonon signals provides an accurate energy estimator for these electron recoils, expressed in keV.

\subsection{Measured spectrum and interpretation}

The data set and fiducial electron recoil spectra used in this study are almost identical to those published in~\cite{Armengaud:2016aoz}. Here, we provide only a brief description of these data, and highlight the differences with respect to~\cite{Armengaud:2016aoz}, which consist in improvements brought to the electron recoil energy estimator. A total exposure of 1149 kg$\cdot$days is selected from WIMP-search data recorded between July 2014 and April 2015. In this selection, 19 detectors are used and have an online threshold of 2~keV or lower. From this 19-detector exposure, a subselection of 289 kg$\cdot$days from 10 detectors is also used with a maximal threshold of 0.8~keV.

Electron recoil spectra are built from a simple set of cuts based on the reconstructed event signals, with respect to their associated baseline resolutions $\sigma_{\rm bl}$. On the one hand, selected events have a fiducial ionization signal $E_{\rm fid}>3.5\sigma_{\rm bl}$. On the other hand, we require that their signals on nonfiducial electrodes be compatible with zero after a cross-talk correction is applied. Multiple-hit events are rejected, based on the data from the complete 24-detector array.

The signals from individual channels are calibrated based on the (8.98, 9.66, 10.37)~keV triplet produced by the cosmogenic activation of $^{65}$Zn, $^{68}$Ge and $^{68}$Ga, as well as on calibration lines from a $^{133}$Ba source. The gains and nonlinearities of phonon channels are calibrated as in~\cite{Armengaud2017c}. In this study, we also correct for the nonlinearity of the ADC used for the ionization measurements. This effect has a 2\% amplitude at 356~keV, and is well described up to 1.4~MeV by a quadratic term that was obtained from the cosmogenic and $^{133}$Ba data, as well as the positions of the peaks from U/Th contaminants.

For fiducial electron recoils at low energy, the optimal energy scale EOP is a linear combination of heat and fiducial energies with relative weights computed from individual baseline noises as in~\cite{Armengaud:2016aoz}. However, above a few tens of keV the energy resolution is severely degraded by charge trapping which spreads the reconstructed ionization signals~\cite{TheEDELWEISSCollaboration2016}, as well as the phonon signals due to the Neganov-Luke effect. We use the procedure described in~\cite{TheEDELWEISSCollaboration2016} to correct trapping effects on the fiducial electrode signals by exploiting charge conservation and the residual signals observed on the veto electrodes. The resulting corrected ionization signal is labeled $E_{\rm fid}^c$. In order to optimize the resolution up to energies of a few hundreds of keV, we therefore use the following empirical energy scale:

\begin{equation}
E = w \,E_{\rm fid}^c + (1-w) \,{\rm EOP}\; ; \; w^{1/2}\equiv 1-e^{-({\rm EOP}/25\,{\rm keV})^2}.
\end{equation}

By fitting the peak positions from radioactive and calibration lines, we checked that the accuracy of this energy scale is better than 0.4\% for all energies considered. Again due to trapping the energy dependence of the resolution for this estimator is of the form $\sigma^2 = \sigma_0^2 + (\alpha\,E)^2$, where $\alpha=1.23$\% for the stacked 19-detector data set and $\sigma_0 = 193$~eV (157~eV) for the 19-detector (10-detector) data set.

 \begin{figure}
 \includegraphics[width=0.5\textwidth,clip=true]{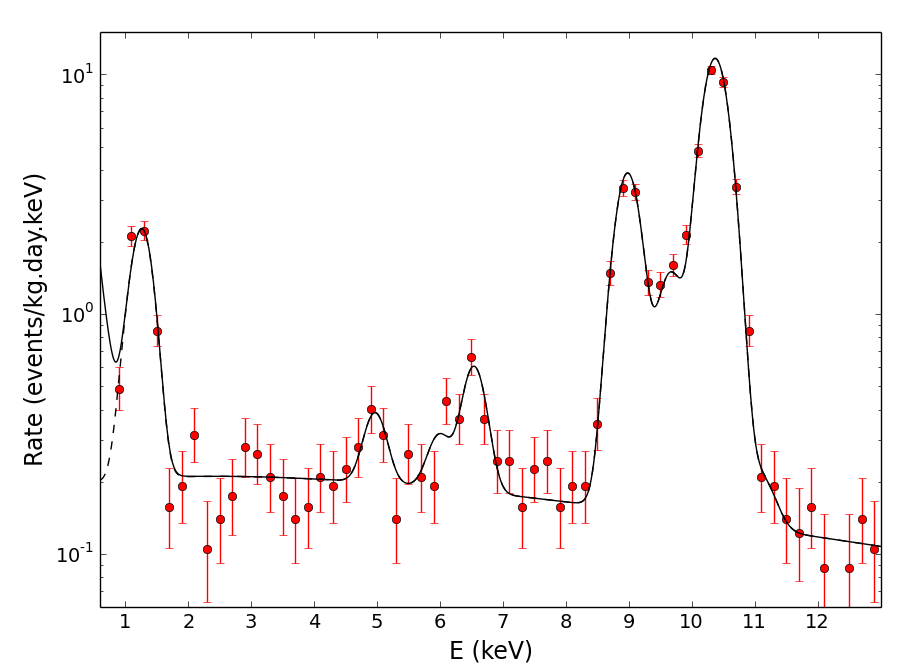}
 \caption{Low-energy electron recoil spectrum of the low-threshold 10-detector subset. The continuous (dashed) line represents the best fit model for the total observed rate $R(E)$ [the corrected recoil rate $R_{\rm ER}(E)$], respectively. This is the same data as represented in the inset from Fig.~2 in~\cite{Armengaud:2016aoz}. \label{fig:spectrum_lowthresh}}
 \end{figure}

The low-energy spectrum is modeled in the same way as in~\cite{Armengaud:2016aoz}. A first component is a flat Compton plateau, whose intensity is $\simeq 0.09$~events/kg$\cdot$day$\cdot$keV for the 19-detector selection. Tritium beta radioctivity was also identified in the fiducial volume of detectors, which originates from cosmic activation of germanium during the time when detectors were located at or close to the ground level. The intensity of this background varies significantly from detector to detector, and its integrated rate for the 19-detector set is on average 1.5 events/kg$\cdot$day. The shape of this beta spectrum is modeled as in~\cite{Armengaud:2016aoz},

\begin{equation}
R(E) \propto (E+m_e)\,(Q_{\beta}-E)^2 \left( 1-e^{-\frac{1.47}{\sqrt{E}}} \right)^{-1}.
\end{equation}

\noindent The endpoint $Q_{\beta}$ is 18.6 keV for tritium decay. Finally, cosmic activation lines are identified from $^{49}$V (4.97~keV), $^{54}$Mn (5.99~keV), $^{55}$Fe (6.54~keV), $^{65}$Zn (8.98~keV), $^{68}$Ga (9.66~keV), and $^{68,71}$Ge (10.37~keV). In the low-threshold, 10-detector data set, an unresolved system of L-shell lines at $1.10-1.30$~keV is also observed in association with the K-shell triplet around 10~keV, as it is visible in Fig.~\ref{fig:spectrum_lowthresh}.

For energies below $\sim 1 - 1.5$~keV, the measured spectrum is also contaminated by a residual population of so-called heat-only events ~\cite{Armengaud:2016cvl}. These events of poorly understood origin have no ionization signals except noise, but the intensity of this background at low energy is such that some of them still can pass the $3.5\sigma$ ionization cut. In addition, for energies close to 1~keV an efficiency correction must be taken into account as described in~\cite{Armengaud:2016aoz}. The observed event rate is therefore modeled by

\begin{equation}
R(E) = \epsilon(E)\times R_{\rm ER}(E) + R_{\rm HO}(E).
\label{eqn:rate}
\end{equation}

$R_{\rm ER}(E)$ is the corrected fiducial electron recoil rate, $\epsilon(E)$ is the efficiency loss from online trigger and analysis thresholds, and $R_{\rm HO}(E)$ is the leakage from heat-only events. Since the distribution of ionization signals for heat-only events is symmetric around zero, the leakage rate above $3.5\sigma$ can be estimated from the observed events with a fiducial ionization below $-3.5\sigma$. We adjust the energy distribution of these events with an exponential law to model $R_{\rm HO}(E)$.
As is shown in Fig.~\ref{fig:spectrum_lowthresh}, the combined effect of both efficiency and heat-only corrections is only significant below 1 keV.

\begin{figure}
 \includegraphics[width=0.5\textwidth,clip=true]{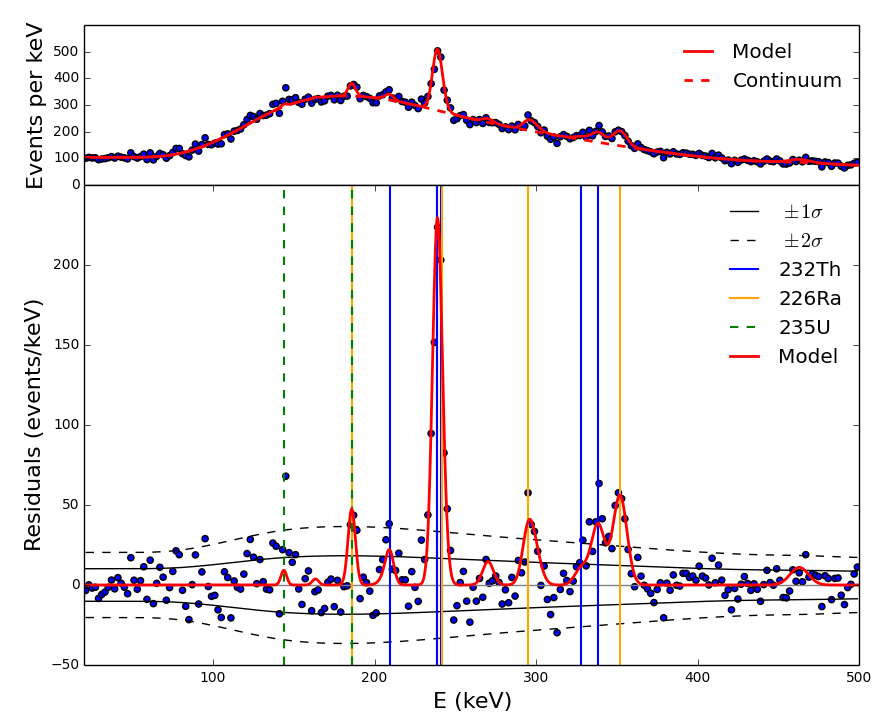}
 \caption{High energy electron recoil spectrum, before (top) and after (bottom) subtracting its continuum component. In the bottom figure, the continuous and dashed envelopes illustrate the 1-$\sigma$ and 2-$\sigma$ statistical fluctuations, respectively. Identified lines from radioactive decay chains are represented.\label{fig:highE_lines}}
 \end{figure}

In the energy range $50\leq E<500$~keV, the background spectrum, presented in Fig.~\ref{fig:highE_lines}, is the sum of a continuum and radioactive lines. Monte Carlo simulations are in good agreement with the shape and intensity of the continuum~\cite{Scorza:2015vla}. However, since we are searching for features localized in a narrow energy range, we prefer to parametrize it using an empirical cubic smoothing spline interpolation.

Radioactive lines were identified in the background data after subtracting this continuum. As illustrated in Fig.~\ref{fig:highE_lines}, all but one features above $\sim 2\sigma$ with respect to statistical fluctuations are identified as lines from the decay chains of $^{232}$Th, $^{226}$Ra and $^{235}$U. Taking into account the relative branching fractions in the decay chains, a simultaneous fit of these different lines yields the rates of 3.0, 1.4 and 0.3 decays per kg$\cdot$day for these three chains. One excess is not fully identified. It is compatible with a line at $E=(145 \pm 0.5)$~keV, well above the energy and expected intensity of the 143.8~keV line from $^{235}$U. Given that its intensity is particularly strong in three detectors located close to each other in the setup, the most probable origin for it is a local source of radioactivity. Possible hypotheses for contaminants as the origin of this line such as $^{141}$Ce or $^{72}$Zn have half-lives that are too short compared to the 10-month period over which the data were recorded.

Given the observed agreement between the data and model over all the energy range considered, upper limits can be placed on the intensity of potential exotic signals.

\subsection{Signal search procedure}

The signatures that are considered are either a monochromatic line at fixed energy, with an intensity $\mu$ expressed in events per kg$\cdot$day; or a CBRD feature with a specific energy dependence as described in the introduction and with an arbitrarily normalized intensity $\mu$. All parameters which describe the background and resolution model presented in the previous subsection form a nuisance vector $\theta$. The exact prescription for $\theta$ as well as for the considered energy range is adapted to each of the searched signatures, so that only the  relevant data and parameters are included. A likelihood function $L(\mu,\theta)$ is defined from the measured binned spectrum, based on Poisson statistics. A Gaussian prior on one of the resolution parameters is also included in $L$ when relevant. Using the notations of~\cite{Cowan2011}, the adopted test statistic is

\begin{equation}
q_{\mu}= \left\{ \begin{array}{ll}
		-2\,{\rm ln}\left( \dfrac{L(\mu,\hat{\hat{\theta}})}{L(\hat{\mu},\hat{\theta})} \right) & \mu > \hat{\mu} \\
		0	& \mu<\hat{\mu} \\
			\end{array}
		\right.
\end{equation}

Here $(\hat{\mu},\hat{\theta})$ is the overall set of signal and parameters which maximizes $L$, while $\hat{\hat{\theta}}$ maximizes $L$ for a fixed value of $\mu$. In all fits we force the signal strength $\mu > 0$. In order to derive 90\% C.L. bounds on the signal strengths $\mu$ as well as sensitivity estimations in the case of absence of signals, we compute the distribution functions for the test statistics $q_{\mu}$, $f(q_{\mu}|\mu)$, and $f(q_{\mu}|0)$ by using appropriate asymptotic formulas as provided in~\cite{Cowan2011}. Based on extensive simulations which take into account all background uncertainties, we checked explicitly that these formulas are highly accurate as long as the searched signal is not a line located on top of an already known radioactive line. In that particular case, the degeneracy between $\mu$ and the radioactive line strength $\theta_{\rm line}$ breaks down the underlying assumptions. From the simulations, we find that the use of the asymptotic formula generates a $\sim 10$\% systematic uncertainty on the signal strength upper bound, which shrinks as soon as the energy difference between the searched and radioactive lines is larger than the local energy resolution. We therefore do not take into account this source of uncertainty when computing upper bounds.

Finally to protect against downward fluctuations of the background we use the CL$_{\rm s}$ prescription~\cite{Read2002}, so that the quoted 90\% C.L. upper bounds $\mu_{\rm up}$ are defined by

\begin{equation}
1-F(q_{\mu} | \mu)_{\rm up} = 0.1\times \left(1-F(q_{\mu}|0)_{\rm up} \right).
\end{equation}

Here $F$ is the cumulative distribution function of the test statistics. This prescription has a conservative coverage, so that the bounds presented here are always at or above $90$\% C.L.

\section{Results}

\subsection{Electron decay}

We start with a search for electrons decaying to invisible particles, a charge conservation violating process considered theoretically e.g.\ in~\cite{OKUN1978597,0038-5670-32-6-R04}. If K-shell electrons in the germanium crystal decay according to $e \rightarrow \nu \nu \nu$, the appearance of a hole in this atomic shell would produce an 11.1~keV X ray. As a particular case of the generic line search with the 19-detector set, our data yields the 90\% C.L. bound $\mu < 3.8\times 10^{-2}$ events / kg$\cdot$day for such a line. This translates into the following bound for the lifetime associated with the process $e \rightarrow$~invisible:

\begin{equation}
\tau_e > 1.2\times 10^{24}\,{\rm years}\;\;\;\text{at 90\% C.L.}
\end{equation}

This is the same numerical value as that obtained by~\cite{Abgrall2016}, as EDELWEISS-III and the MAJORANA demonstrator have very similar performance (energy resolution and background) at 11.1~keV.

\subsection{Solar axions}

 \begin{figure}
 \includegraphics[width=0.5\textwidth,clip=true]{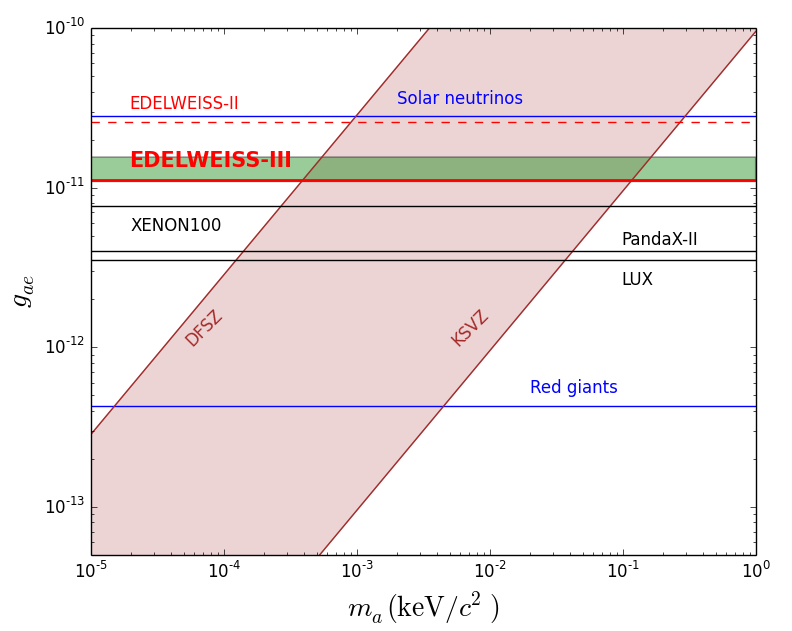}
 \caption{Limit on the axion-electron coupling $g_{ae}$ obtained from the lack of observation of a CBRD signal. The green band represents the $2\sigma$ expected sensitivity. Other limits obtained from similar searches are shown in black (\cite{Akerib2017,Fu2017,Aprile2014}). The blue lines show indirect bounds from the measured solar neutrino flux~\cite{PhysRevD.79.107301} and from the observed tip of the red-giant branch~\cite{PhysRevLett.111.231301}. \label{fig:cbrd}}
 \end{figure}

Since the CBRD signal is broadly peaked around 2~keV, we combine both the 19-detector, 2-keV threshold dataset and the additional 10-detector subset in the 0.8 -- 2~keV energy range together in the likelihood function. Due to downward fluctuations visible in both independent spectra, the derived upper limit is smaller than the average expectation. The probability to obtain a limit lower than the measured one is estimated to be 4\%. The bound on the arbitrary signal strength $\mu$ translates into

\begin{equation}
g_{ae} < 1.1\times 10^{-11}\;\;\;\text{at 90\% C.L.}
\end{equation}

This value, together with the expected sensitivity, is compared in Fig.~\ref{fig:cbrd} to both axion models and other experimental bounds. This search excludes QCD axion models with masses $m_a>0.39$~eV/c$^2$ (DFSZ) or $m_a>118$~eV/c$^2$ (KSVZ scenario). This is the best bound obtained so far with germanium detectors. While better sensitivities are obtained by dual-phase Xenon TPCs, germanium detectors would have the capability to provide a spectroscopic confirmation of a potential signal.

 \begin{figure}
 \includegraphics[width=0.5\textwidth,clip=true]{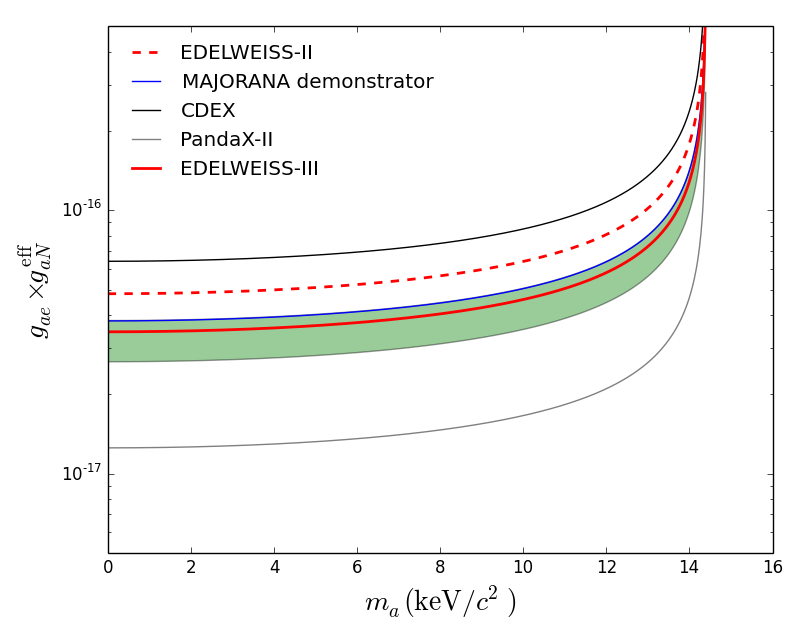}
 \caption{90\% C.L. bounds on the combination $g_{ae}\times g_{aN}^{\rm eff}$ as a function of axion mass, derived from the absence of observation of a 14.4~keV line. The green band represents the $1\sigma$ expected sensitivity. Other limits from similar searches are also shown~\cite{Armengaud2013b,Abgrall2016,Liu2017,Fu2017}. \label{fig:iron}}
 \end{figure}

As a particular case of the generic line search, the analysis of the 19-detector data set yields a bound $\mu < 2.05\times 10^{-2}$~events / kg$\cdot$day for a potential line at 14.4~keV. This translates, for axions with a mass $\ll 14$~keV/c$^2$, into the following bound:

\begin{equation}
 g_{ae}\times g_{aN}^{\rm eff} < 3.5\times 10^{-17} \;\;\;\text{at 90\% C.L.}
 \end{equation}

The variation of this bound as a function of $m_a$ is shown in Fig.~\ref{fig:iron}, together with similar constraints from other experiments. In particular, we notice that, as for the case of electron decay, EDELWEISS-III and the MAJORANA demonstrator yield nearly identical results. As for the CBRD channel, experiments using Xenon detectors such as PandaX-II have a better sensitivity, although germanium-based experiments would provide a clearer spectral identification if a signal were observed. In specific QCD axion frameworks, the bound on the 14.4~keV intensity alone excludes the mass ranges $6.6\,{\rm eV} < m_a < 14.4$~keV/c$^2$ for DFSZ axions, and $130\,{\rm eV}<m_a < 14.4$~keV/c$^2$ in the KSVZ scenario.

\subsection{Bosonic dark matter absorption}

 \begin{figure*}
 \includegraphics[width=0.49\textwidth,clip=true]{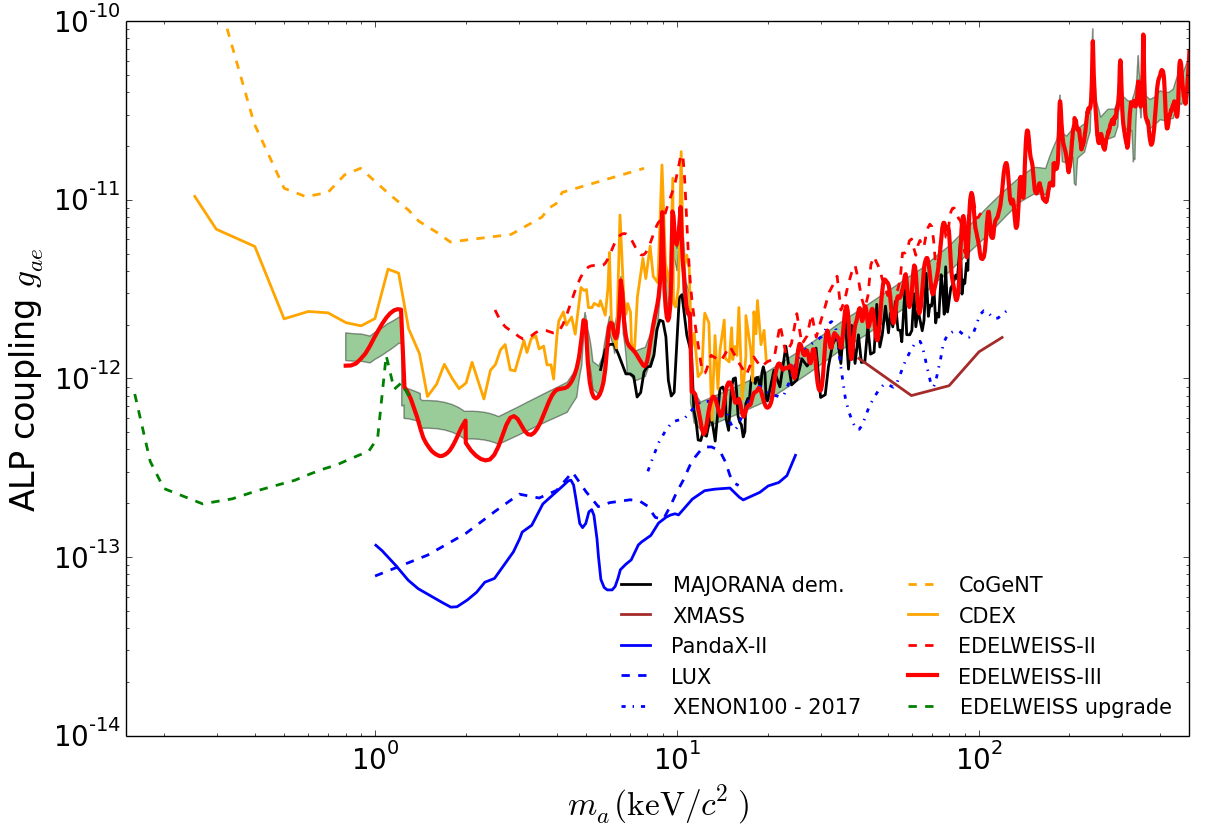}
 \includegraphics[width=0.49\textwidth,clip=true]{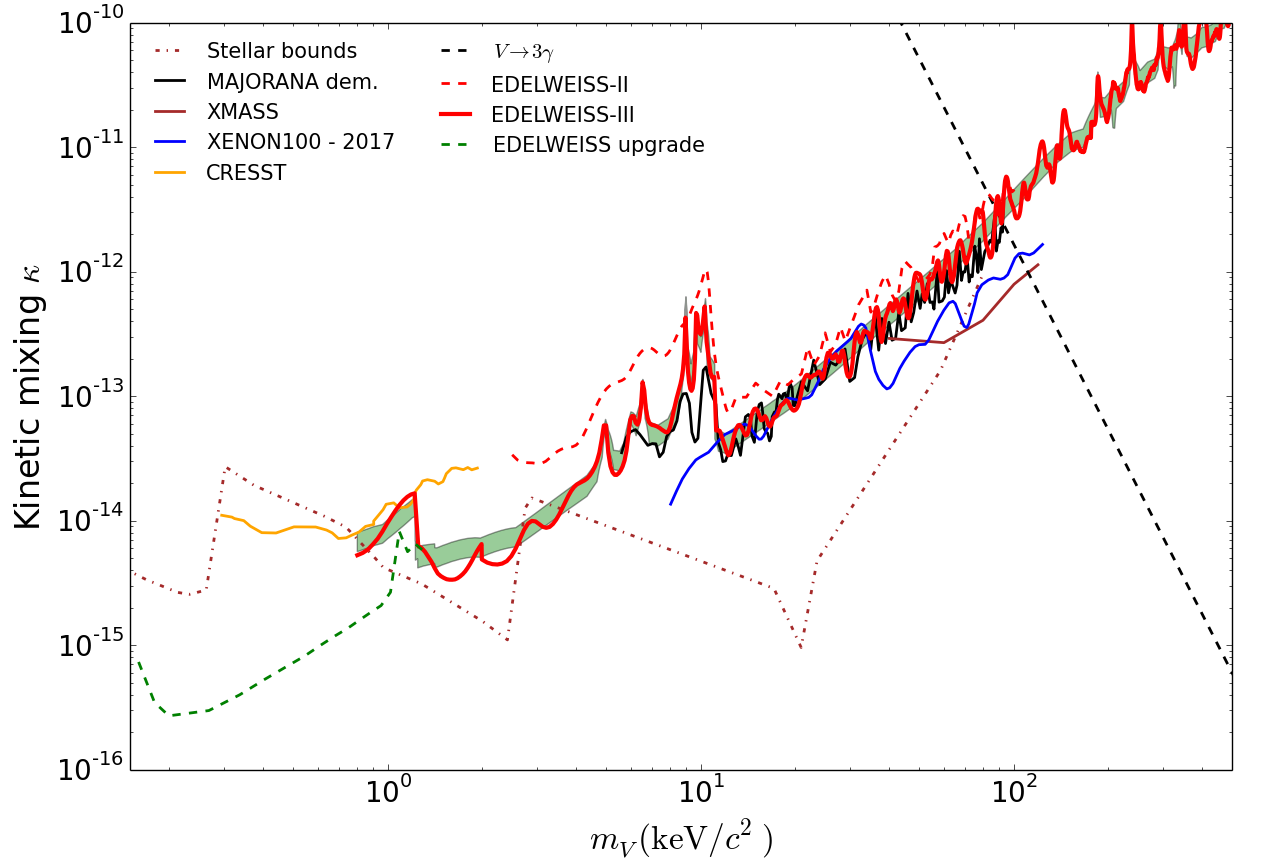}
 \caption{Left: Bounds on the ALP dark matter coupling to electrons $g_{ae}$ from EDELWEISS-III and other direct search experiments (\cite{Abgrall2016,Abe2014,XENONcollaboration2017,Armengaud2013b,Akerib2017,Fu2017,Liu2017,Aalseth:2008rx}). Right: Bounds on the hidden photon DM kinetic mixing $\kappa$ from direct searches, including~\cite{Angloher:2016rji,Armengaud2013b,XENONcollaboration2017,Abgrall2016,Abe2014}. Astrophysical bounds from~\cite{An:2014twa} are also included for the hidden photon scenario. The green bands show the $1\sigma$ sensitivity for EDELWEISS-III. The green dashed lines are sensitivity projections for upgraded EDELWEISS detectors.\label{fig:dm_bounds}}
 \end{figure*}

The search procedure described in Sec. II-C is finally used to set an upper limit on a generic line feature in the 0.8 -- 500~keV energy range. The data sets, energy range of the fits, energy binning choices, and background models were adapted as a function of each of the scanned energies. In particular, the 19-detector data set is used for line searches above 2~keV, while the 10-detector subset is used for line searches in the 0.8 -- 2~keV range. The derived upper bounds on line strength $\mu$ are then converted into bounds on the DM couplings. Figure~\ref{fig:dm_bounds} presents the bounds on $g_{ae}$ (left) for the ALP scenario, and on $\kappa$ (right) for the case of a hidden photon. 

The constraints set over a wide mass range are competitive with other searches, and yield a significant improvement with respect to EDELWEISS-II. We also extended the range of explored bosonic dark matter masses up to 500~keV/c$^2$, although other astrophysical probes become particularly constraining in many scenarios above 100~keV/c$^2$ --- as illustrated by the $V \rightarrow 3\gamma$ bound in the case of vector DM, shown in Fig.~\ref{fig:dm_bounds} (right). The EDELWEISS-III sensitivity and bounds are at the same level as those from the MAJORANA demonstrator in the 6 -- 100~keV/c$^2$ mass range. In the 1 -- 6~keV/c$^2$ mass range, we provide the best constraints on bosonic dark matter couplings from a spectroscopic germanium experiment, and also extend the range of masses probed down to 800~eV/c$^2$, below the threshold of Xenon-based detectors. Remarkably, for hidden photons with masses smaller than $\sim 900$~eV/c$^2$ our bounds and sensitivity to the kinetic mixing $\kappa$ are at the same level of the stellar bounds derived from anomalous energy loss in horizontal branch (HB) stars~\cite{An:2014twa}.

\section{Conclusion and prospects with upgraded EDELWEISS detectors}

From the measurement of electron recoils in the fiducial volume of EDELWEISS-III detectors, we have derived constraints on several hypothetical processes beyond the Standard Model, namely the emission of axions or ALPs from the Sun, the absorption of bosonic keV-scale dark matter particles from our galactic halo, and the electron decay to three neutrinos. These bounds represent a significant improvement with respect to previous EDELWEISS-II results. For processes with an associated electron recoil energy larger than 6~keV, the sensitivity and bounds presented here are similar to those recently published by the MAJORANA demonstrator experiment. In the case of processes with deposited energies $\le 6$~keV, we provide the best limits from a spectroscopic germanium-based experiment, and start to explore new parameter space for bosonic DM scenarios with a mass below 1~keV.

In the near future, an upgrade of EDELWEISS detectors will permit to extend the mass reach of bosonic DM searches well below 800~eV. The 800~eV energy threshold from this data set is indeed driven by the readout noise of ionization channels, and by the phonon sensitivity of the absorber-NTD system. Within the same infra\-structure, we plan to make use of high electron mobility transistors~\cite{Phipps:2016mwv}, ideally suited for the readout of low-capacitance interleaved electrodes, to read charge signals replacing the existing JFETs. The reduction in the readout noise will permit us to reject the background from interactions taking place near the surface of the detectors down to 50~eV energy deposits, and obtain a 35~eV RMS resolution on the ionization signal for fiducial interactions. To improve the resolution in the search of lines in the electron recoil spectrum, the heat signal can be boosted by applying a bias of 20~V, a value at which it was tested that the interleaved ring electrodes of our 870 g units are operating properly. With this boost, the 300~eV phonon resolution obtained on present-day detectors can translate into a 40~eV rms resolution for electron recoils. By making use of both phonon and ionization signals, a combined electron recoil energy resolution of 25~eV rms is expected. The clear identification of fiducial electron recoils will be possible above 100 -- 150~eV.

Based on these performances, we estimate the sensitivity to bosonic dark matter absorption for a DM mass down to 150~eV/c$^2$, with a 500 kg$\cdot$day exposure and assuming the same background levels as measured with EDELWEISS-III. This sensitivity is shown in Fig.~\ref{fig:dm_bounds} (dashed green lines). It will permit to explore unknown territory in parameter space in the mass range 0.1 -- 1~keV. It will exceed stellar cooling bounds from the Sun or HB stars in the case of hidden photons~\cite{An:2014twa}. In the case of ALPs, it will approach the region of parameter space ($g_{ae}\lesssim 2\times 10^{-13}$) which is consistent with an ALP explanation for the white dwarf luminosity function~\cite{Bertolami:2014wua}.

\begin{acknowledgments}

The help of the technical staff of the Laboratoire Souterrain de Modane and the participant 
laboratories is gratefully acknowledged. The EDELWEISS project is supported in part by the 
German Helmholtz Alliance for Astroparticle Physics (HAP), by the French Agence Nationale 
pour la Recherche (ANR) and the LabEx Lyon Institute of Origins (ANR-10-LABX-0066) of 
the Universit\'e de Lyon within the program ``Investissements d'Avenir'' (ANR-11-IDEX-00007), 
by the P2IO LabEx (ANR-10-LABX-0038) in the framework ``Investissements d'Avenir'' 
(ANR-11-IDEX-0003-01) managed by the ANR (France), by Science and Technology Facilities 
Council (UK), and the Russian Foundation for Basic Research (Grant No. 18-02-00159).

\end{acknowledgments}

\bibliography{axions_edw3}

\end{document}